\documentclass[]{interact}

\usepackage{epstopdf}
\usepackage[caption=false]{subfig}
\usepackage{color}
\usepackage{amssymb}
\usepackage{lipsum}
\usepackage{graphicx}
\usepackage{xcolor}
\usepackage{siunitx}

\usepackage[numbers,sort&compress]{natbib}
\bibpunct[, ]{[}{]}{,}{n}{,}{,}
\renewcommand\bibfont{\fontsize{10}{12}\selectfont}
\makeatletter
\def\NAT@def@citea{\def\@citea{\NAT@separator}}
\makeatother

\theoremstyle{plain}

\theoremstyle{definition}

\theoremstyle{remark}

\begin{document}

\articletype{ARTICLE}

\title{\textit{Investigating the Molecular Design Mechanism Behind the Hydrophobicity of Biological Surface Nanostructures: Insights from Butterfly and Mosquito Systems}}

\author{
\name{Fan Meng\textsuperscript{a} and Noriyoshi Arai\textsuperscript{a}$^{\ast}$\thanks{$^\ast$Corresponding author. Email: arai@mech.keio.ac.jp; Fax: +81 45 566 1495; Tel: +81 45 566 1846}}
\affil{\textsuperscript{a}Department of Mechanical Engineering, Keio University, Yokohama, Japan.}
}

\maketitle

\begin{abstract}
Wettability is a fundamental physicochemical property of solid surfaces, with unique wettability patterns playing pivotal roles across diverse domains. Inspired by nature's ingenious designs, bio-inspired materials have emerged as a frontier of scientific inquiry. They showcase remarkable hydrophobic properties observed in phenomena such as mosquitoes preventing fog condensation, and lotus leaves exhibiting self-cleaning attributes.

This groundbreaking research delves into the hydrophobic characteristics of biomimetic surfaces using coarse-grained molecular simulation and the free energy barrier evaluation system. By analyzing the butterfly wings and mosquito eyes model, we aim to pioneer a comprehensive framework that factors in the influence of surface parameters on the free energy barrier. Through meticulous simulation and analysis, we strive to validate and enhance the reliability of the free energy barrier assessment method, deepening our understanding of hydrophobicity across diverse biomaterials and paving the way for optimizing their properties for a myriad of applications.

During our investigation, we shed light on the elusive intermediate state, a departure from the typical Cassie or Wenzel state, enriching our theoretical framework for surfaces with distinctive properties. This research is a catalyst for developing biomimetic materials with superior hydrophobic characteristics and innovative fabrication processes, transcending academic boundaries and promising significant strides in environmental conservation, medicine, and beyond, offering hope for a greener, healthier, and more sustainable future.
\end{abstract}


\section{Introduction}
\label{introduction}
Wettability stands as a fundamental physicochemical property characterizing solid surfaces, with surfaces exhibiting unique wettability patterns proving particularly captivating. Such surfaces play pivotal roles across diverse domains, including energy utilization, environmental protection, healthcare, sustainable development, and factory manufacturing. Exploring extreme wettability commenced with observing specialized phenomena in nature, prompting detailed investigations into the underlying mechanisms \cite{DARMANIN2015, Bhushan2009, koch2008, Barthlott1997}. Drawing inspiration from nature's ingenuity, bio-inspired materials have emerged as a frontier of scientific inquiry, offering boundless possibilities across various disciplines. For instance, mosquitoes possess the ability to prevent fog condensation on their compound eyes in humid environments \cite{SONG2017, GAO2007}, lotus leaves exhibit self-cleaning attributes owing to their slender downy hairs \cite{Bhushan2006, Barthlott1997}, water striders navigate surfaces at high speeds due to their super-draining structure \cite{Gao2004}, and fish evade sticking to oil underwater \cite{Liu2009}. Even butterfly wings remain unsoaked amidst rain \cite{Zheng2007}. Nature presents an array of templates featuring superhydrophobic structures, showcasing remarkable hydrophobic properties.

In the context of self-cleaning surfaces, routine maintenance often necessitates sanitizing materials to uphold cleanliness. Especially during the COVID-19 pandemic, widespread reliance on disinfection products has escalated, exacerbating the environmental burden posed by disinfectant residues \cite{Dhama2021, Sivakumar2021, Nagendra2020, Samara2020} Conversely, in nature, myriad surfaces exhibit inherent self-cleaning capabilities, mitigating the need for external interventions. The application of self-cleaning technology is ubiquitous, spanning from window glass to solar cell panels and flat panel displays\cite{Min2008, Kavitha2020}. This inherent mechanism, characterized by robust self-cleaning and superhydrophobicity prowess independent of external assistance, has garnered significant research interest.\cite{Byun2009, Zhang2008, Solga2007, Barthlott2016, Bhushan2011, Liu2012}.

In our past study, we have endeavored to delve into the hydrophobicity of surface materials by evaluating free energy barriers. Given the vast array of biomaterials and their multifaceted structural characteristics, our exploration thus far has focused on mosquitoes' compound eye structure. While this provided valuable insights, it represents merely a fraction of the rich diversity phenomenon in nature. In this study, we aim to expand our investigative vision by encompassing the hydrophobic properties of butterfly wings, which are renowned in natural sciences for their classical hydrophobic structure. We intend to extend our analysis to these intricate and fascinating structures by incorporating the free energy barrier evaluation system.

By meticulously examining biological structures such as butterfly wings and mosquito eyes, we endeavor to harness the potential of the free energy barrier assessment method to evaluate their hydrophobic properties precisely. Alka Jaggessar and colleagues have proposed that surface parameters, including size, width, spacing, tip sharpness, and aspect ratio, significantly influence the antimicrobial efficiency of surfaces\cite{Jaggessar2017}. 
By further evaluating and simulating additional biomimetic material surfaces using the same method, Our objective is to rigorously test the accuracy and reliability of the free energy barrier assessment method. Through this expanded analysis, encompassing a broader range of biomimetic materials and surface structures, we seek to ascertain the method's robustness and applicability in biomaterials research. This iterative approach will allow us to validate the efficacy of the assessment method and gain deeper insights into the underlying principles governing hydrophobicity across diverse biomaterials. Ultimately, such validation efforts are essential for establishing the method as a reliable tool for evaluating and optimizing biomaterial properties, paving the way for its broader adoption in both research and industrial applications.

By amalgamating theoretical validation with empirical results, we aspire to enhance the efficiency of assessing and optimizing the hydrophobic properties of biomaterials. This endeavor holds promise for advancing our fundamental understanding of biological systems and translating these insights into industrial applications. From biomimetic materials with enhanced hydrophobic characteristics to innovative fabrication processes, the potential ramifications of our research are far-reaching, spanning domains such as environmental conservation, healthcare, and beyond.

\section{Methods and Model}
 \subsection{Many-body dissipative particle dynamics}
We employed the many-body dissipative particle dynamics (MDPD) method\cite{Warren2001, Warren2003, Zhao2021} to investigate how the nanostructured solid surface relates to the free energy barrier.
In the classical dissipative particle dynamics (DPD) method\cite{Hoogerbrugge92, Espanol1995, groot97}, the interaction between particles only involves repulsive forces that reflect the average of the conservative forces between several coarse-grained molecules\cite{Kinjo2007, Kinjo2007_2}.
Therefore, the classical DPD method, such as gas-liquid interfaces, cannot reproduce sharp density differences.
To address this limitation, the DPD conservation forces have incorporated appealing terms, including:
\begin{equation}
\mathbf{F}_{ij}^\mathrm{C} = a_{ij} \left( 1 - \frac{\left|\mathbf{r}_{ij}\right|}{r{\mathrm{c}}} \right) \mathbf{n}_{ij} + b_{ij} \left( \bar{\rho}_i + \bar{\rho}_j \right) \left( 1 - \frac{\left|\mathbf{r}_{ij}\right|}{r{\mathrm{d}}} \right) \mathbf{n}_{ij}.
\label{eq:FC}
\end{equation}
Here, $\mathbf{r}$ is the position vector, $\mathbf{r}_{ij} = \mathbf{r}_{j} - \mathbf{r}_{i}$, $\mathbf{n}_{ij} = \frac{\mathbf{r}_{ij}}{\left|\mathbf{r}_{ij}\right|}$, $\bar{\rho}_i$ represents the local density at the particle, and $r{\mathrm{c}}$ and $r_{\mathrm{d}}$ are cutoff distances used to determine the effective range of the force.
The initial term represents an attractive interaction, while the second term accounts for the many-body effect and acts as a repulsive interaction.
As a result, the values of $a_{ij}$ and $b_{ij}$ are selected to be negative and positive, respectively.
The local density is calculated using the following formula:
\begin{equation}
\bar{\rho}_i = \sum_{i \neq j} \frac{15}{2 \pi r_{\mathrm{d}}^3} \left( 1 - \frac{\left|\mathbf{r}_{ij}\right|}{r{\mathrm{d}}} \right)^2.
\label{eq:ek}
\end{equation}
The reports currently utilize reduced units for the cutoff radius $r_{\mathrm{c}}$, the particle mass $m$, and the energy $k_\mathrm{B}T$. Here, $T$ denotes the temperature, and $k_\mathrm{B}$ represents the Boltzmann constant.
Thus, $r_{\mathrm{c}} = m = k_\mathrm{B}T = 1.0$, and the time unit is defined as $\tau = \frac{\sqrt{m r_{\mathrm{c}}^2}}{k_\mathrm{B} T} = 1.0$.

\subsection{Simulation Model}
\begin{figure*}
	\centering 
	\includegraphics[width=14cm]{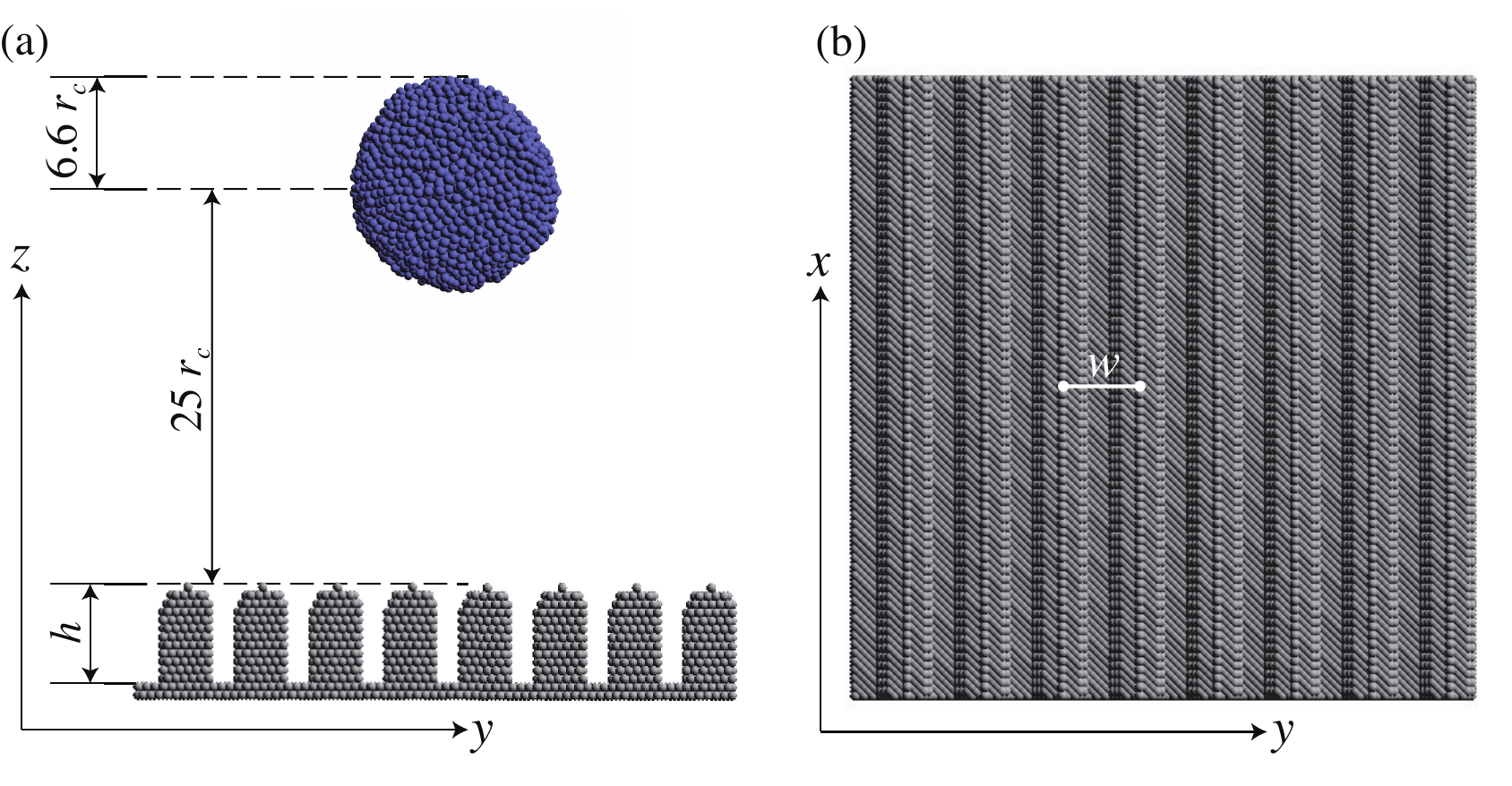}	
	\caption{(a) The initial configuration encompasses the droplet and the solid surface. The droplet, characterized by a radius of $r_\mathrm{droplet}$, assumes a size of 6.6 $r_c$. Positioned atop the nano-bump, the droplet maintains a separation distance of 25 times $r_c$ from the bump's apex. The height of the bump is denoted as $h$.
(b) The top view of the model surface, inspired by the butterfly wings.The spatial arrangement of the nano-bumps is defined by the center distance $w$, contributing to the overall pattern on the surface.} 
	\label{initially model}%
\end{figure*}

\begin{figure*}
	\centering 
	\includegraphics[width=14cm]{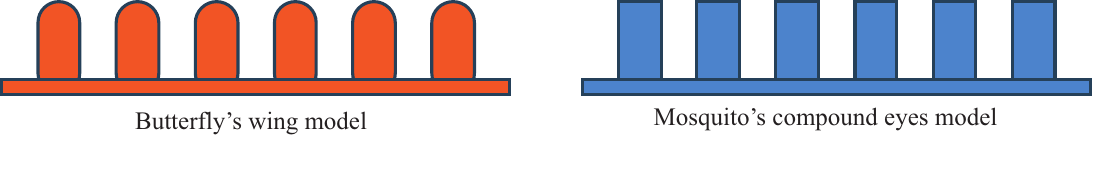}	
	\caption{The side profile diagrams of the butterfly wing model and the mosquito compound eye model.} 
	\label{modelimage}
\end{figure*}

The investigation involved modeling the bio-inspired superhydrophobic structure of butterfly wings to explore how the spacing and shape of nano-structures affect the size of the free energy barrier Figure \ref{initially model}. In contrast to previous models \cite{Meng2023}, which featured equidistant alternating nano-pillars on the surface, our current model comprises protruding nano-bumps and grooves equally distributed on the surface. Additionally, as depicted in Figure \ref{modelimage}, the summit of the surface roughness structure in the butterfly wing model differs from that of the mosquito compound eye model. The apex of the nanostructures on the butterfly wing model exhibits a rounded shape, whereas that of the mosquito compound eye model appears flat. We denote the distance between each group of nano-bumps as $w$, and the height of the nano-bumps as $h$.

\begin{figure*}
	\centering 
	\includegraphics[width=14cm]{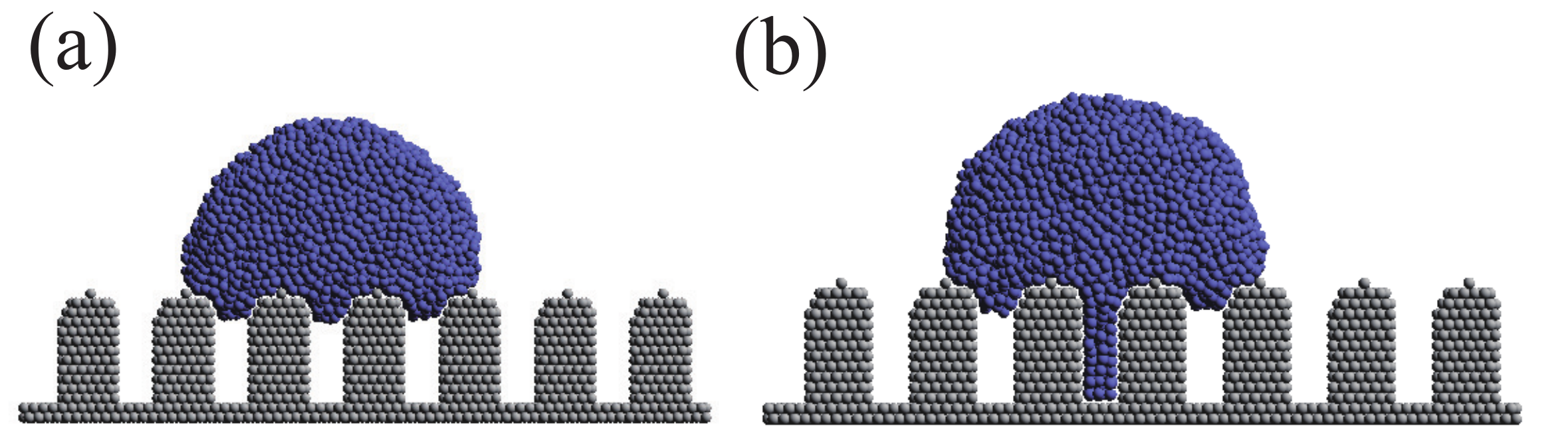}	
	\caption{The states of droplets on the surface: (a) Cassie state, (b) Wenzel state} 
	\label{Cassie_Wenzel}%
\end{figure*}

A water droplet with a radius of 6.6 $r_\mathrm{c}$ was positioned at 25 $r_\mathrm{c}$ from the top of the nano-bump. This droplet was given kinetic energy and directed downward to impact the nano-surface. Upon impact, the droplet exhibited either a Cassie state or Wenzel state\cite{Cassie, Wenzel}, shown in Figure \ref{Cassie_Wenzel} with the transition being a probabilistic event. Although higher kinetic energy favored the Wenzel state, it was not guaranteed. However, when the kinetic energy exceeded a critical threshold, the droplet consistently transitioned to the Wenzel state.

Concerning the wall models, we constructed ten systems of two types characterized by varying particle numbers ($N$): 73,272; 84,144; 93,072; 93,432; 102,720. One type investigated the correlation between the free energy barrier and $w$, achieved by altering the center-to-center distance of the bumps ($w$), while the other type examined the relationship with $h$, manipulated by adjusting the height of the bumps ($h$). Different wall parameter configurations corresponded to distinct particle numbers, as delineated in Table \ref{sym_conditions}.

\begin{table}[h]
\small
  \caption{\ Simulation conditions for each surface system.}  
  \begin{tabular*}{0.45\textwidth}{@{\extracolsep{\fill}}llllllll}
    \hline
    Num & $w$ & $h$ & $L_x$ & $L_y$ & $L_z$ & $N$\\
    \hline\hline
   \multicolumn{7}{c}{Group of changed $w$} \\ \hline
   1 &  450 & 6 & 36 & 36 & 60 & 102720 \\
   2 &  500 & 6 & 36 & 36 & 60 & 93432  \\
   3 &  525 & 6 & 36 & 36 & 60 & 93072 \\
   4 &  600 & 6 & 36 & 36 & 60 & 84144 \\
   5 &  660 & 6 & 36 & 36 & 60 & 73272 \\
    \hline
       \multicolumn{7}{c}{Group of changed h} \\ \hline
   6 & 525 & 4 & 36 & 36 & 60 & 68880 \\
   7 & 525 & 5 & 36 & 36 & 60 & 80976 \\
   8 & 525 & 6 & 36 & 36 & 60 & 93072 \\
   9 & 525 & 7 & 36 & 36 & 60 & 105168 \\
   10 & 525 & 8 & 36 & 36 & 60 & 117264 \\
   \hline\hline
   \label{sym_conditions}
  \end{tabular*}
\end{table}

All simulations were conducted in a constant volume and constant temperature ensemble. The thermostat was implemented using pairwise dissipative and random forces coupled through the fluctuation-dissipation theorem. Here, $\sigma$ and $\gamma$ represent dissipative and random forces, respectively, satisfying $\sigma = \sqrt{2} \gamma k_\mathrm{B} T$, thereby reproducing a canonical ensemble. We fixed $\sigma$ and $\gamma$ at 3.0 and 4.5, respectively, while maintaining the temperature at 0.5 $k_\mathrm{B}T$ to mitigate the impact of thermal fluctuations.

In our simulations, particles in the solvent are denoted as S. In contrast, particles in the wall are denoted as W. Drawing from our previous investigation \cite{Kadoya2017}, the interaction parameters in Equation (1) are specified as $a_\mathrm{SS} = -40 k_\mathrm{B} T$, $a_\mathrm{SW} = -25 k_\mathrm{B} T$, and $b_\mathrm{SS} = b_\mathrm{SW} = 25 k_\mathrm{B} T$. We set the cutoff radius $r_\mathrm{d}$ for the repulsive conservative force at 0.75 $r_\mathrm{c}$ and the time step $\Delta t$ at 0.005 $\tau$. Each simulation run extended throughout 5,000 $\tau$ to ensure stable droplet formation.


\section{Result}
\subsection{Measurement of free energy barrier}
\label{calculate}
Under each simulation system and for every $e_\mathrm{k}$(kinetic energy) value, the surface undergoes 54 impacts. After each of these impacts, outcomes are meticulously examined, and the quantities of Cassie and Wenzel states are tallied. We can precisely determine the Wenzel probability corresponding to varying $e_\mathrm{k}$ values under fixed surface parameters through robust statistical analysis. Subsequently, by substituting both the Wenzel probability$(P_\mathrm{w})$ and its corresponding $e_\mathrm{k}$ into equation \ref{eq_DG}:

\begin{equation}
\label{eq_DG}
P_\mathrm{w} = P_{0} \exp (-\frac {\Delta G_\mathrm{cw}} {e_\mathrm{k}})
\end{equation}

Here, $P_0$ represents the pre-exponential factor, $\Delta G_\mathrm{cw}$ is defined as the free-energy barrier from the Cassie to Wenzel state, and $e_\mathrm{k}$ corresponds to the kinetic energy of the droplet's center of mass.

\begin{equation}
\label{eq_ek}
{e_\mathrm{k}} = \frac{1}{2} m v_z^2.
\end{equation}
$P_0$ is derived through the equation:
\begin{equation}
\label{eq_P0}
\left(\frac{e_{k1}}{e_{k2}} - 1\right) \ln(P_0) = \frac{e_{k1}}{e_{k2}} \ln(P_{w1}) - \ln(P_{w2})
\end{equation}

The magnitude of the free energy barrier for each distinct set of surface parameters is ascertained. This methodological approach comprehensively elucidates the intricate relationship between $e_\mathrm{k}$, surface properties, and the ensuing free energy barriers.

For the precise computation of the free energy barrier within this simulation, meticulous selection of appropriate kinetic energy values was essential. Careful kinetic energy assignment was undertaken to ensure the droplet achieved an optimal Wenzel probability after impacting the nano-surface. This precision was vital for accurate calculation employing formula \ref{eq_DG}. Specifically, we selected five distinct $w$ values for surface parameters and five corresponding $e_\mathrm{k}$ values for impact simulations at each $w$ value. 54 impact simulations were conducted for each $e_\mathrm{k}$, resulting in 1350 simulations across all systems. The MDPD time step was fixed at 100, with a total simulation step count of 15,000 for each group.

\begin{figure}
	\centering 
	\includegraphics[width=8cm]{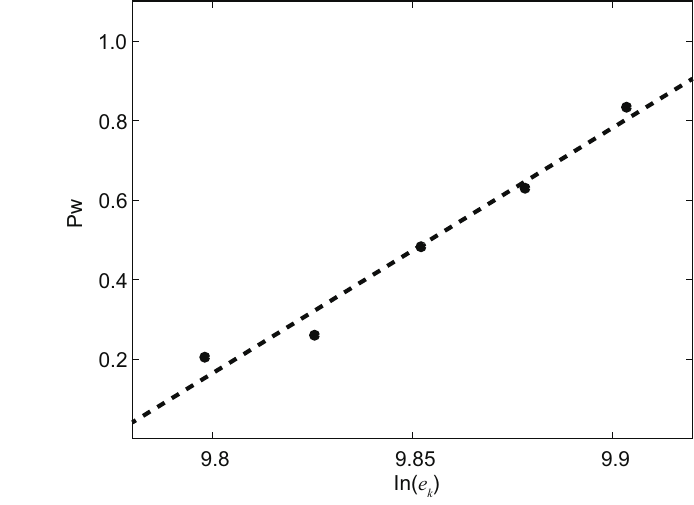}	
	\caption{The $y$-axis represents the probability of the Wenzel state ($Pw$), while the $x$-axis corresponds to the natural logarithm of the kinetic energy ($e_\mathrm{k}$).} 
	\label{pw}
\end{figure}

As Fig. \ref{pw} shows, for $w = 5.25$ and $h = 6.00$, the simulation results revealed that the probability of Wenzel state ($p_w$)  formation varied with the initial kinetic energy values of the droplet: 18,000, 18,500, 98,000, 19,500, and 20,000, showing probabilities of 20.4 \%, 25.9 \%, 48.1 \%, 63.0 \%, and 83.3 \%, respectively.
The free energy barrier can be calculated once $e_\mathrm{k}$ and $p_w$ are obtained.

\subsection{Effect on the free energy barrier at bump distances ($w$)}
\label{fitting}
\begin{figure}
	\centering
	\includegraphics[width=8cm]{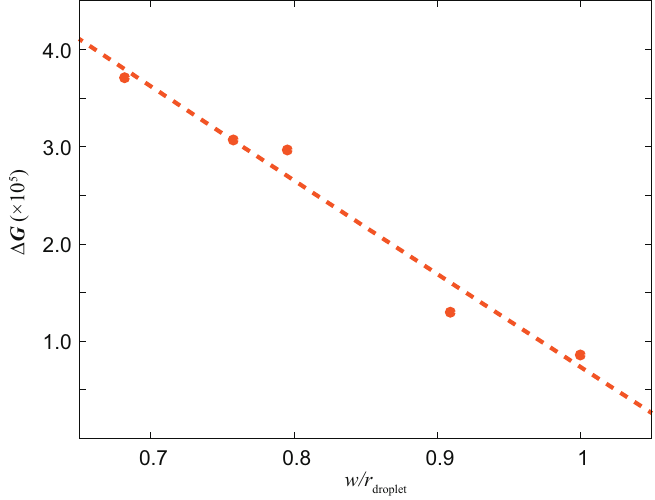}
	\caption{Effect of different $w$ values on $\Delta G$.}
	\label{wset}
\end{figure}

In our initial series of simulations, we explored the relationship between the parameter $w$ and the corresponding free energy barrier. We specifically investigated five different ratios of the nano-bump distance $w$ to the droplet radius $r_{\text{drop}}$: 0.68, 0.76, 0.80, 0.91, and 1.00. Using the methodology detailed in Section 5, we determined the magnitude of the free energy barriers associated with various surface parameters. These findings were then fitted to quadratic equations of the form $\Delta G = \mathrm{a}x^{2} + \mathrm{b}x + \mathrm{c}$, utilizing the constants $a = 2.94376 \times 10^4$, $b = -1.0123 \times 10^6$, and $c = 1.05623 \times 10^6$. The results, along with a plot of the fitted equations, are presented in Fig.\ref{wset} Although we employed a quadratic function for fitting, the resulting curve closely resembles a linear function.

From the results, it is evident that the size of the free energy barrier increases gradually as the ratio of $w$ to $r_{\text{drop}}$ decreases. This indicates a higher hydrophobicity of the material, suggesting a strong relationship between the spacing of the nano-bumps and the size of the free energy barrier (hydrophilicity of the material).

\subsection{Effect on the free energy barrier at bump height ($h$)}
\begin{figure}
	\centering 
	\includegraphics[width=8cm]{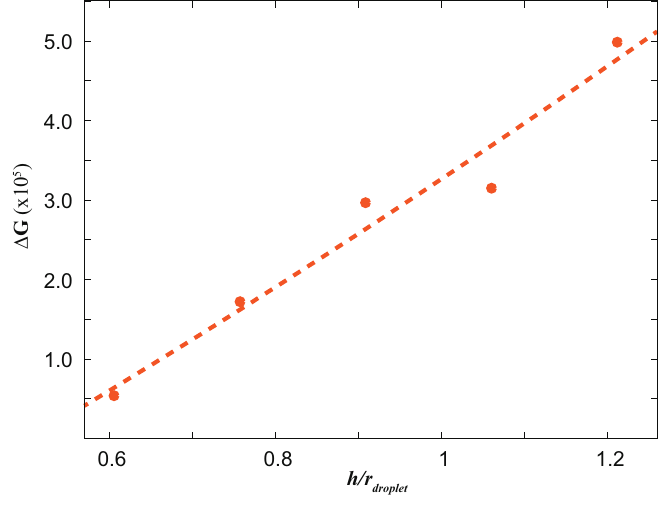}	
	\caption{The effect of different $h$ values on $\Delta G$.} 
	\label{hset}
\end{figure}

In the second series of simulations, we investigated the impact of varying the height of the surface nano-bump, represented as $h$, on the magnitude of the free energy barrier. We employed five distinct ratios of $h$ to $r_{\text{drop}}$: 0.61, 0.76, 0.91, 1.06, and 1.21. Using the same analytical approach described in Section 5, we synthesized the outcomes of these simulations. We fitted them to an equation of the form $\Delta G = ax^2 + bx + c$, where $a = -1.7541 \times 10^6$, $b = 4.12036 \times 10^6$, and $c = -1.92904\times 10^6$ represent constants. The results are illustrated in Figure \ref{hset}.

From the data presented in Figure \ref{hset}, it is evident that the free energy barrier at the surface gradually increases with the height of the nano-bump. Despite fitting the results to a quadratic curve, the final fitted curve closely follows the linear trend observed in the data.

\section{Discussion}
\begin{figure*}
	\centering 
	\includegraphics[width=14cm]{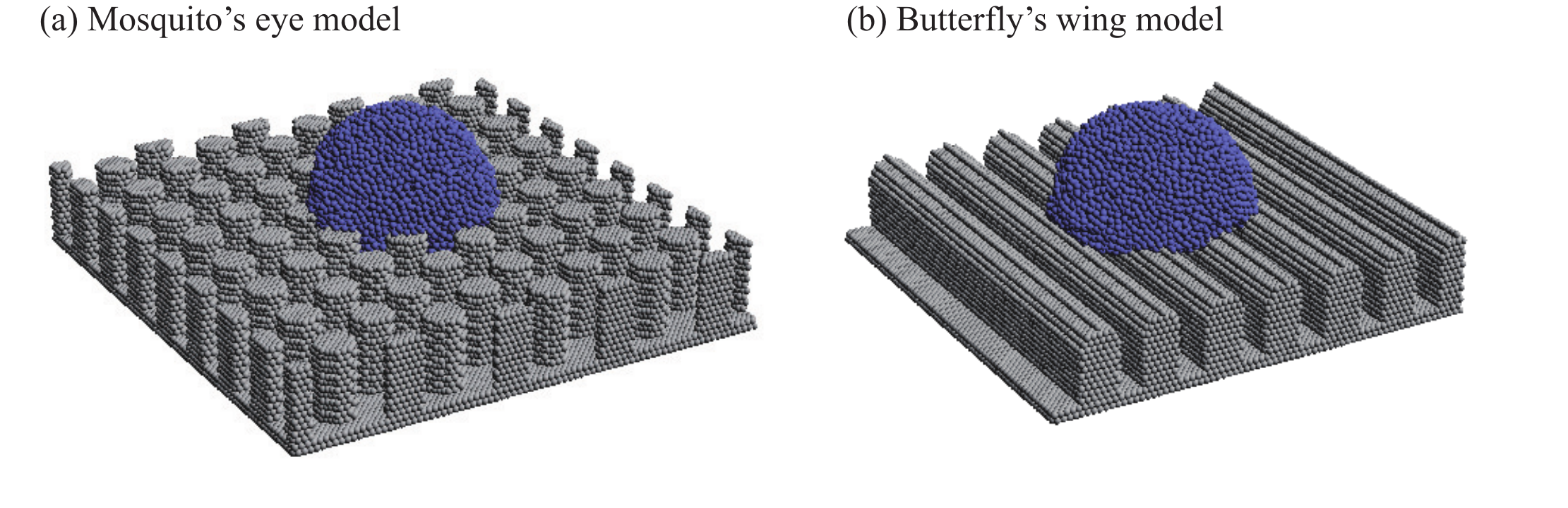}	
	\caption{Rough surface structure of mosquito eye model and butterfly wing model.(a)Mosquito's eye. (b)Butterfly's wing}
	\label{modelcompare}
\end{figure*}

We compared the curves depicting changes in the free energy barriers of butterfly wings with those of a model representing the compound eye of a mosquito. The surface roughness of both the mosquito and butterfly models is illustrated in Figure\ref{modelcompare}. The left part (a) of the figure presents the model of the mosquito's compound eye characterized by alternating surface hexagonal nanopillars, while the right part (b) illustrates the model of the butterfly's wing, featuring nano bumps distributed periodically at specific intervals. This comparative analysis of two distinct surface nanostructured biomimetic materials provides valuable insights into the intricate relationship between surface roughness and free energy barriers.

From the figure\ref{wset}, it is evident that as $w$ decreases—indicating a reduction in the distance between the bumps—the free energy barrier of the surface increases. A diminishing distance ($w$) between the nanobumps on both sides enhances the likelihood of nanobumps supporting droplets upon surface impact. Consequently, it becomes challenging for the droplet to transition into a Wenzel state after surface impact.

\begin{figure}
	\centering 
	\includegraphics[width=8cm]{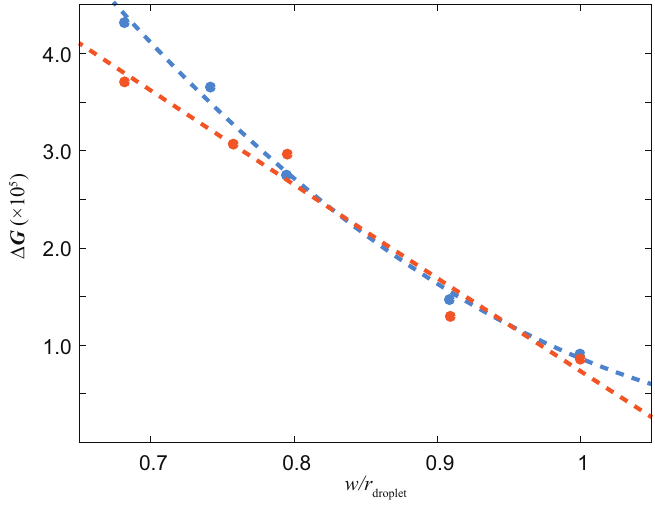}	
	\caption{Comparison of the free energy barrier curve ($\Delta G$) of the mosquito eye model with the free energy barrier curve ($\Delta G$) of the butterfly wing is illustrated in the graph. The mosquito's eye data is depicted in blue, while the butterfly's wing data is represented in orange.}
	\label{wcompare}
\end{figure}

The results of our novel model simulation of the butterfly's wing closely align with the trend observed in the free energy barrier curve from the preceding simulation of the mosquito's eye model, shown in Fig.\ref{wcompare}. One of the unique surface parameter variables investigated in this study is the ratio of the nano-bump to the droplet radius.
    
A decrease in this ratio indicates a narrowing spacing of the nanobumps. Conversely, considering the ratio from another perspective signifies a widening gap between the droplet and the nanobumps. Notably, natural nanostructures found in organisms such as mosquito eyes, butterfly wings, and lotus leaf surfaces are inherently fixed; they cannot arbitrarily adjust the height and spacing of these nanostructures. Consequently, in this study, the ratio of the nanobump's pitch $ w $ to the droplet's radius $ r_{\text{drop}} $ is set to 1, representing an already diminutive droplet size in nature. Specifically, 6.6$ r_{\text{c}} $ equals 5.60 nanometers.
    
Based on the outcomes of this simulation, it is evident that both the nanostructure of the mosquito's eye and the butterfly's wing demonstrate the characteristic wherein a smaller $ w/r_{\text{drop}} $ value correlates with increased hydrophobicity. This implies that the mosquito's eye and the butterfly's wing exhibit greater hydrophobicity towards larger water droplets.

\begin{figure}
\centering 
\includegraphics[width=8cm]{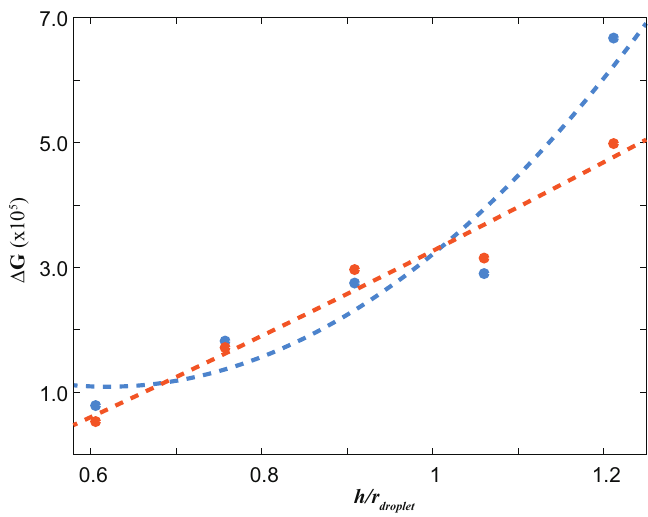}	
\caption{The effect of different values on $\Delta G$.} 
\label{hsetcompare}
\end{figure}

As evident from Fig.\ref{hsetcompare}, mosquito eyes and butterfly wings models demonstrate a consistent trend: the increase in surface nanobump height (h) leads to a corresponding rise in hydrophobic free energy barriers. This observation highlights the pronounced influence of nanobump height on the hydrophobicity of surfaces. As nanobump height escalates, the constraints on interactions between surface molecules intensify, consequently enhancing surface hydrophobicity. Such insights are pivotal for understanding surface properties relevant to hydrophobic interactions, particularly in surface chemistry and biomaterial design contexts. Further exploration of this relationship holds promise for informing strategies to tailor surface hydrophobicity for diverse applications, ranging from self-cleaning surfaces to biomedical implants\cite{Jaggessar2017, Kavitha2020}.

The w-$\Delta G$ and h-$\Delta G$ curves of the mosquito eye and butterfly wing models exhibit similar trends.   Despite the contrasting surface roughness patterns, the influence of final surface parameters on the free energy barrier consistently follows a comparable trajectory. While the surface structures differ, the effects of h and w on the free energy barrier exhibit a similar influence.   The distinct surface structures are attributed to the differing functionalities of mosquito eyes and butterfly wings in nature.  

Mosquito eyes necessitate hydrophobicity and antireflectivity to facilitate ample light transmission for an expansive field of view while concurrently upholding self-cleaning capabilities\cite{SONG2017}. Conversely, butterfly wings must possess hydrophobic, self-cleaning properties and optical properties\cite{Zhang2015, Wagner1996}. Hence, despite disparate objectives, both organisms have evolved a distinctive hydrophobic surface architecture over extended periods of evolution to fulfill their respective functional requisites.

\begin{figure*}
\centering
\includegraphics[width=13cm]{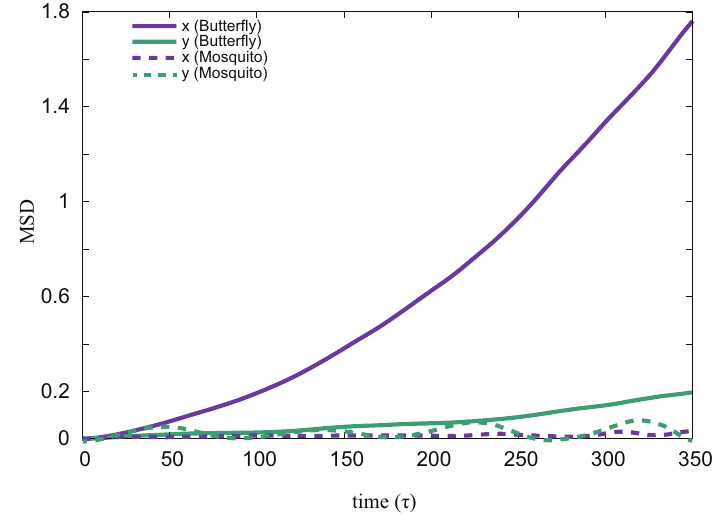}
\caption{MSD about Butterfly and Mosquito}
\label{MSD}
\end{figure*}

Since butterfly wings exhibit distinctive anisotropic flow characteristics\cite{Bixler2012, Bixler2013}, we conducted Mean Square Displacement (MSD) analysis on our results, as illustrated in Figure \ref{MSD}. In the butterfly wing model, it is evident from the MSD plots that the displacement of droplets along the $x$-axis, parallel to the wing grooves, is significantly greater than that along the $y$-axis, vertical to the grooves. This observation strongly suggests that droplets are more prone to sliding motion along the $x$-axis direction. By contrast, a notable discrepancy emerges when comparing this analysis with the MSD results obtained from mosquito models. Specifically, the MSD values for droplets along the $x$- and $y$-directions in the mosquito model are nearly identical, indicating the absence of any discernible anisotropic behavior in the mosquito model.

Yoshimitsu $et$ $al.$ also pointed out that by comparing how water droplets slid on surfaces with either pillar or groove structures\cite{Yoshimitsu2002}. They found that the surface with grooves, where the water touched a larger solid area, had a lower water contact angle than the pillar surface(The magnitude of the free energy barrier is smaller); this observation aligns closely with our simulation findings. Interestingly, the groove surface showed better water shedding in the parallel direction. This suggests that creating the right three-phase line is more effective than increasing contact angles by reducing the solid-water interface\cite{Oner2000}. These findings reinforce our simulation results, providing further evidence for our study. 
We again corroborate this fact at the more microscopic molecular design level and gain a deeper understanding of it.

\begin{equation}
\label{free energy 3d}
\Delta G^{2} (w,h)= \Delta G (w) \cdot \Delta G (h)
\end{equation}
Here, $\Delta G(w) = \mathrm{a}w^{2} + \mathrm{b}w + \mathrm{c}$, and $\Delta G(h) = \mathrm{d}h^{2} + \mathrm{e}h + \mathrm{f}$, a, b, c, d, e, f are fitting parameters.  

We endeavored to integrate the functions $w$-$G$ and $h$-$G$ into a three-dimensional coordinate system depending on the function \ref{free energy 3d}, with the $x$- and $y$-axes representing the ratios of $w$ and $h$ to $r_\mathrm{droplet}$, and the $z$-axis representing $\Delta G^{2}$. 

\begin{figure*}
\centering
\includegraphics[width=13cm]{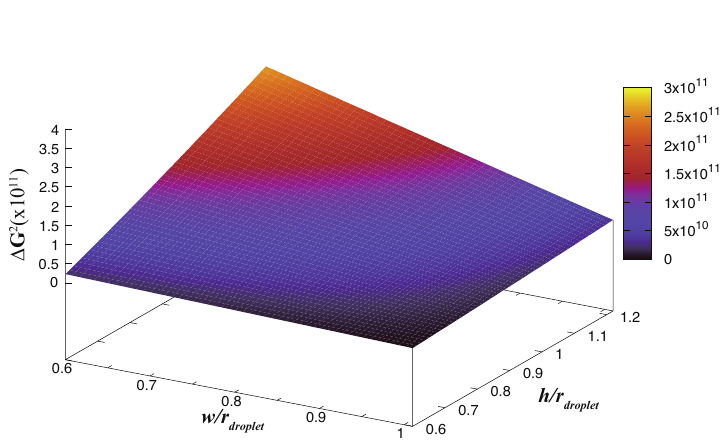}
\caption{Integrate the $w$-$G$ and $h$-$G$ functions into a three-dimensional coordinate system}
\label{3d}
\end{figure*}

By this methodology, we can integrate the outcomes of this study into three-dimensional (3D) equations, enabling a more intuitive examination of how changes in both w and h affect the magnitude of free energy barriers. As depicted in the figure\ref{3d}, as the ratio $ w/r_{\text{drop}} $ decreases and $ h/r_{\text{drop}} $ increases, the free energy barrier attains its maximum value. This observation aligns with our previous findings in the mosquito eye model\cite{Meng2023}, thus reaffirming this pattern. However, it is noteworthy that unlike the abrupt increase observed in the mosquito eye model, the rate of increase in the free energy barrier demonstrates an overall linear trend. Furthermore, the 3D function representation indicates that the surface trend is more similar to a 3D flat surface; it differs from the 3D curved surface when we integrate it into the previous 3D representation.

On a separate note, it is essential to highlight that the influence of varying w on the free energy barrier becomes less apparent when h is tiny. This finding further supports our earlier assertion: in the industrial production of superhydrophobic surfaces, adjusting the height of surface roughness ($h$) holds incredible promise in achieving superhydrophobic surfaces compared to modifying the distance between surface roughness features ($w$).

In our last study, there was an intermediate state after w/r or h/r exceeded a specific value. This intermediate state does not belong to the typical Cassie or Wenzel state. In this study, we still encounter this intermediate state, but the last column structure is less frequent than the frequency of the intermediate state. \"Oner's study noted that the surface exhibits superhydrophobicity when the radius of the column is 2-32\(\mu\)m and the distance between the columns is similar, and when the radius of the central column is 64-128\(\mu\)m and the distance between the columns is similar, the water droplets invade the column between the columns and are trapped between the columns, and are fixed on the surface. The scale we simulate is relatively small; the droplet radius $r_\mathrm{droplet}$ ratio to the surface column width is 2.2 times. Although it is not known precisely what the droplet radius \"Oner used if he used a 1mm(1000\(\mu\)m) radius droplet ratio to the surface column would be about 8-15 times. So this also explains why our simulation had many intermediate states last time. Through our simulation, we can better observe the intermediate phenomenon of water droplets invading between the pillars but not touching the bottom. Due to the limitations of current observational instruments and other technologies, the current explanation for this phenomenon is that there is no better way to use molecular simulation technology. The explanation of this phenomenon provides a great theoretical reference for surfaces with specific properties.

\section{Summary and conclusions}
In summary, our comparative analysis of the free energy barriers of butterfly wings and mosquito compound eyes sheds light on the relationship between surface roughness and hydrophobicity. Through surfaces with different nanostructured patterns, we observed that decreasing the spacing between nano-bumps ($w$) increased the free energy barrier. This decrease in spacing enhanced the likelihood of nano-bumps supporting droplets upon surface impact, making it challenging for the droplet to transition into a Wenzel state. Additionally, increasing the height of surface nano-bumps ($h$) corresponded to a rise in hydrophobic free energy barriers, indicating the pronounced influence of nanobump height on surface hydrophobicity.

Furthermore, despite differences in surface roughness patterns between butterfly wings and mosquito compound eyes, both demonstrated a similar trend in the influence of $w$ and $h$ on the free energy barrier. This suggests that while the surface structures differ, the effects of $w$ and $h$ on the free energy barrier exhibit a consistent influence.

Our findings contribute to a deeper understanding of surface properties relevant to hydrophobic interactions, with implications for applications such as self-cleaning surfaces and biomedical implants. Overall, our study emphasizes the significance of surface roughness parameters in dictating free energy barriers and underscores the potential of adjusting roughness to customize surface properties for specific applications. Furthermore, the newfound understanding and insights into intermediate states offer fresh perspectives for crafting specific hydrophobic surface structures. 

On the other hand, this study has only explored the relationship between surface parameters $w$ and $h$ and the free energy barrier. The design of three-phase lines (including shape, length, width, continuity of contact, and amount of contact) is also crucial and cannot be overlooked. In the future, we will also consider incorporating it into our evaluation system.

\footnotesize{
\bibliography{ref1}
\bibliographystyle{gMOS} 
}

\clearpage

\end{document}